\documentclass[aps,prl,twocolumn,floatfix,superscriptaddress,showpacs,amsmath,amssymb]{revtex4-1}
\usepackage{graphicx}
\usepackage{epsfig}
\usepackage{braket}
\allowdisplaybreaks

\newcommand{\be}{\begin{equation}}
\newcommand{\ee}{\end{equation}}
\newcommand{\bea}{\begin{eqnarray}}
\newcommand{\eea}{\end{eqnarray}}
\newcommand{\ba}{\begin{eqnarray*}}
\newcommand{\ea}{\end{eqnarray*}}

\newcommand{\bk}{\mathbf{k}}

\newcommand{\bp}{\mathbf{p}}

\newcommand{\up}{\uparrow}
\newcommand{\down}{\downarrow}

\newcommand{\eps}{\varepsilon}

\begin{document}

\title{Transient Dynamics of $d$-wave Superconductors after a Sudden Excitation}
\author{Francesco Peronaci}
\affiliation{International School for Advanced Studies (SISSA/ISAS) and CNR-IOM Democritos, Via Bonomea 265, 34136, Trieste, Italy}
\author{Marco Schir\'o}
\affiliation{Institut de Physique Th\'{e}orique, Universit\'{e} Paris Saclay, CNRS, CEA, F-91191 Gif-sur-Yvette, France}
\author{Massimo Capone}
\affiliation{International School for Advanced Studies (SISSA/ISAS) and CNR-IOM Democritos, Via Bonomea 265, 34136, Trieste, Italy}


\begin{abstract}
 Motivated by recent ultrafast pump-probe experiments on high-temperature superconductors, we discuss the transient dynamics of a $d$-wave BCS model after a quantum quench of the interaction parameter.
 We find that the existence of gap nodes, with the associated nodal quasiparticles, introduces a decay channel which makes the dynamics much faster than in the conventional $s$-wave model.
 For every value of the quench parameter, the superconducting gap rapidly converges to a stationary value smaller than the one at equilibrium.
 Using a sudden approximation for the gap dynamics, we find an analytical expression for the reduction of spectral weight close to the nodes, which is in qualitative agreement with recent experiments.
\end{abstract}

\date{\today}
\maketitle

\emph{Introduction. - } Recent advances in time-resolved spectroscopies have triggered a growing interest in the transient dynamical behavior of high-temperature superconductors optically excited far from equilibrium.
 By shining the sample with intense ultrafast pulses (pump) one can trigger nonequilibrium transient states, whose physical properties are then recorded by a second pulse (probe) which hits the sample with a given time delay.
 This approach opens up a wealth of information unavailable to conventional time-averaged spectroscopies~\cite{Giannetti_NatComm11,DalConteEtAlScience12,HintonEtAlPRL13,Carbone,NovelliEtAlNatureComms,DalConteEtAlNaturePhys,CoslovichEtAlPRL13,CilentoEtAlNatureComms,Perfetti_PRL15}.
 When used in combination with angle-resolved photoemission spectroscopy (ARPES), these ultrafast methods allow us to track and follow in real time the evolution of quasiparticle modes in different momentum sectors, such as those close to the gap nodes and antinodes of $d$-wave cuprate superconductors~\cite{GrafEtAlNatPhys11,SmallwoodEtAlScience12,SmallwoodEtAlPRB14}.
 In addition, when irradiation is sufficiently strong and optically tailored in such a way to selectively excite specific modes, one can even stabilize transient states with fundamentally different physical properties~\cite{Ichikawa_NatMat11,Fausti_Science11,Caviglia_PRL12}.
 A spectacular example is given by recent experiments reporting signatures of light-induced metastable superconductivity in cuprates at much higher temperature than at equilibrium~\cite{Kaiser_PRB14,HuEtAl_NatMat14}.

 These experimental breakthroughs raise a number of intriguing questions.
 From one side, a nontrivial and rich transient dynamical behavior is expected in correlated materials, which feature complex phase diagrams characterized by competing phases~\cite{PA_Lee_Cuprates}.
 On a more fundamental level, pump-probe experiments suggest the possibility to explore novel metastable phases that can only be accessed along nonthermal pathways, e.g., by means of photoexcitation. 
 Theoretical investigations along these directions have appeared in the literature in recent years~\cite{Kollar_prb08,Freericks_prl09,Moritz_prb10,Kollar_prb09} addressing questions like the thermalization of pump-excited Mott or Kondo insulators~\cite{Eckstein_prb11,EcksteinWernerPRB12} and the role of lattice vibrations~\cite{Phonons_Schiro_PRB2012}, orbital degrees of freedom~\cite{SandriFabrizio_PRB15} and competing orders~\cite{FuHungSachdevPRB14,MoorEtAlPRB14} in the relaxation dynamics.
 The dynamics of conventional superconductors has also been studied with reference to pump-probe experiments~\cite{Papenkort_PRB07,Papenkort_PRB08,Zachmann_NJP13} or in the presence of electron-phonon interactions~\cite{SchnyderManskeAvellaPRB11,KrullEtAlPRB14,KemperEtAl_arxiv14} and Coulomb repulsion~\cite{MazzaFabrizioPRB12}.
 In the context of ultracold atomic systems, the research has focused on the $s$-wave weak coupling BCS regime~\cite{BarankovLevitovSpivakPRL04,YuzbashyanEtAlPRB05,BarankovLevitovPRL06,YuzbashyanEtAlPRL06}, with recent attempts to extend the analysis to the crossover into the BEC regime~\cite{GurariePRL09,YuzbashyanEtAlPRA15} and to exotic order-parameter symmetries~\cite{FosterEtAlPRB13,FosterEtAlPRL14}.

 Yet, despite this recent activity, many fundamental questions concerning specifically the transient response of superconducting materials remain wide open.
 In particular, a characteristic feature of high-$T_c$ cuprates is the momentum anisotropy of the superconducting gap, which has a $d_{x^2-y^2}$ symmetry (which we indicate as $d$-wave in the following).
 This leads to the existence of nodal lines along which the superconducting gap vanishes, a characteristic feature which is sometimes used as a definition of exotic superconductivity.
 These gapless excitations significantly affect the thermodynamics and spectral properties as compared to the conventional $s$-wave case, where the gap is uniform over the whole Fermi surface, and they can possibly play an even bigger role in the nonequilibrium dynamics.

 In this Letter we consider a minimal model of a $d$-wave superconductor and the simplest nonequilibrium protocol of a quantum quench of the interaction parameter.
 Within a BCS-like mean-field approximation, we calculate the gap dynamics and compare the results with those for the conventional $s$-wave superconductor, identifying the distinctive features descending from the existence of the nodes.
 Then, we derive an approximate formula for the spectral weight which is directly relevant to ARPES experiments in $d$-wave superconductors.
 Of course a mean-field description, while reasonable for the deeply overdoped region of the phase diagram of cuprates, is unable to capture the strong correlations which dominate the low-doping region, leading to a complex interplay between different phases.
 However, we believe it is useful to highlight the physics of a simple model with $d$-wave symmetry, as this will help to disentangle the effect of the anisotropic gap in more involved calculations in which competing orders, strong-coupling effects, and any other realistic feature are included.
 
\emph{Model and gap dynamics. - } As a starting point of our analysis we consider the two-dimensional BCS Hamiltonian with a momentum-dependent separable interaction in the $d$-wave channel
 \begin{equation}
 H=\sum_{\bk\sigma}\eps_{\bk}\,c^{\dagger}_{\bk\sigma}\,c_{\bk\sigma}-J\sum_{\bk\,\bp}\,\gamma_{\bk}\gamma_{\bp}\,c^{\dagger}_{\bk\up}\,c^{\dagger}_{-\bk\down}\,c_{-\bp\down}\,c_{\bp\up},
 \label{hamiltonian}
 \end{equation}
 with $\eps_{\bk}=\vert\bk\vert^2-\mu$ and $\gamma_{\bk}=\cos2\theta\sim k_x^2-k_y^2$ where $\theta$ is the polar angle in $k$ space.
 The superconducting gap, or order parameter, is defined as
 \begin{equation}
 \Delta_{\bk}=\Delta\gamma_{\bk},\qquad\Delta=J\sum_\bk\gamma_\bk\braket{c_{-\bk\down}c_{\bk\up}},
 \label{delta}
 \end{equation} 
 and it vanishes along the nodal lines $k_x=\pm k_y$. This leads to modes at arbitrary low energies $E_\bk=\sqrt{\eps_\bk^2+\vert\Delta_{\bk}^2\vert}$ which, as we show below, have important consequences on the relaxation dynamics.

 The existence of the nodal lines marks a fundamental difference not only with the $s$-wave symmetry which has $\gamma_\bk=1$ and a fully gapped spectrum~\cite{YuzbashyanEtAlPRB05,YuzbashyanEtAlPRL06}, but also with unconventional symmetries such as $p+ip$ ($\gamma_\bk\sim e^{i\theta}$)~\cite{FosterEtAlPRB13} and $d+id$ ($\gamma_\bk\sim e^{2i\theta}$)~\cite{MarquetteNPhysB13} where the gap vanishes at most for one point in momentum space. Notably, in all the above nodeless cases the $\theta$ dependence of the gap function can be gauged away leaving us with an effective one-dimensional problem of the Richardson-Gaudin form~\cite{RichardsonShermanNPhysB64,GaudinJphysFrance76} which can be solved exactly.

 In the absence of such a full solution, we resort to a time-dependent BCS variational {\it{ansatz}} $\vert\Psi(t)\rangle=\Pi_{\bk}\left(u_{\bk}(t)+v_{\bk}(t)c^{\dagger}_{\bk\uparrow}c^{\dagger}_{-\bk\downarrow}\right)\vert0\rangle$. This is equivalent to introducing a quasi-particle Hamiltonian
 \begin{equation}
 H_\mathrm{qp}(t)= \sum_{\bk\sigma}\eps_{\bk}\,c^{\dagger}_{\bk\sigma}\,c_{\bk\sigma}-\sum_{\bk}\,\left(\gamma_\bk \Delta(t)c^{\dagger}_{\bk\uparrow}c^{\dagger}_{-\bk\downarrow}+\mathrm{hc}\right)
 \label{eqn:Hqp}
 \end{equation}
 and solving the equations of motion for the expectation values $\braket{c_{\bk\sigma}^\dagger(t)c_{\bk\sigma}(t)}$ and $\braket{c_{-\bk\down}(t)c_{\bk\up}(t)}$ with a time-dependent order parameter $\Delta(t)=J\sum_\bk\gamma_\bk\braket{c_{-\bk\down}(t)c_{\bk\up}(t)}$ which is calculated self-consistently at each time and couples the different points in $k$ space~\cite{SM_dynDwaveSC}.

 To drive the system out of equilibrium we consider the simplest and very popular quantum quench protocol.
 We take at $t=0$ the ground state of the Hamiltonian [Eq.~(\ref{eqn:Hqp})] with a given value of the interaction parameter $J=J_i$ and calculate the time-evolution according to the same Hamiltonian but with a different interaction $J=J_f$.
 It is convenient to discuss the results in terms of $\Delta_i$ and $\Delta_f$, the gaps which the system would have at equilibrium for $J_i$ and $J_f$, respectively.

\begin{figure}[t]
 \begin{center}
 \begin{tabular}{ccc}
 \includegraphics[width=40mm]{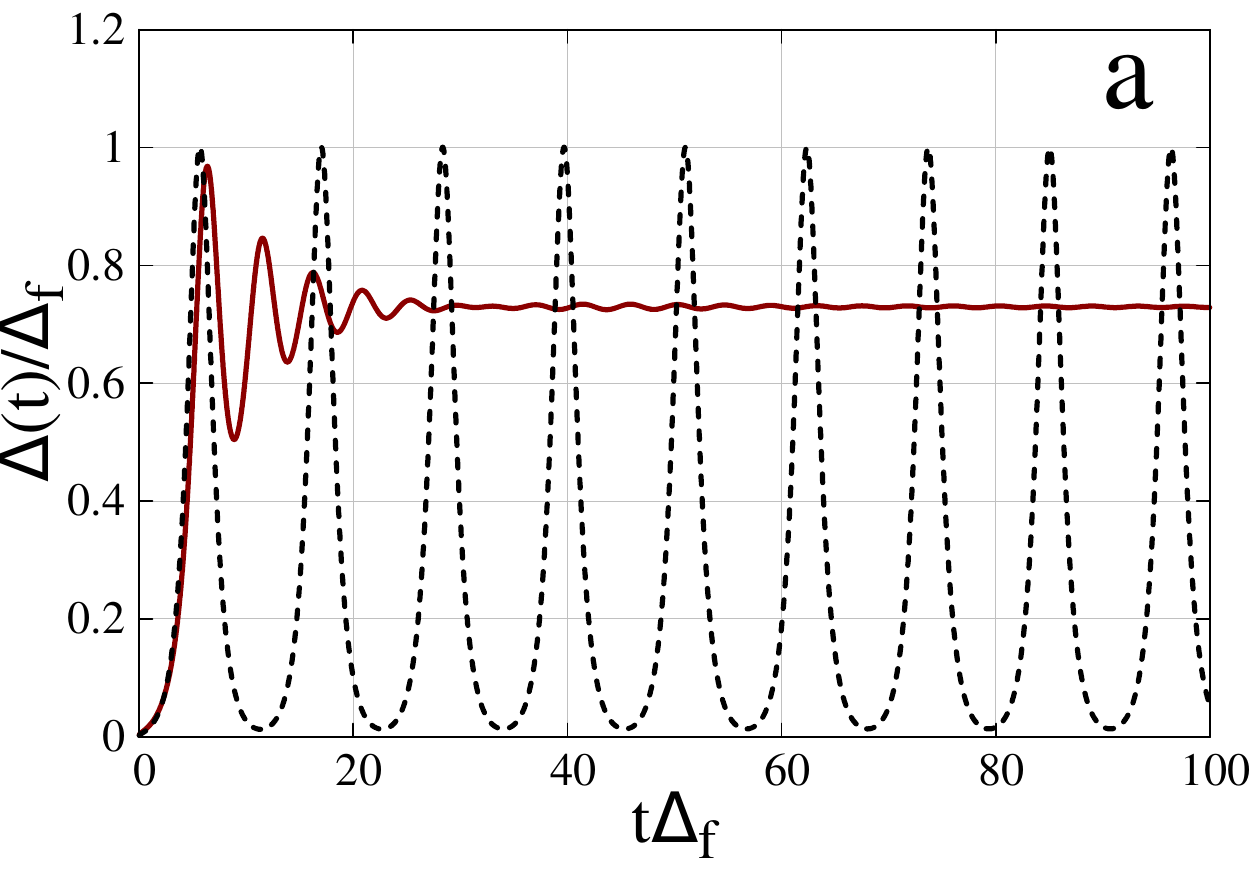}&
 \includegraphics[width=40mm]{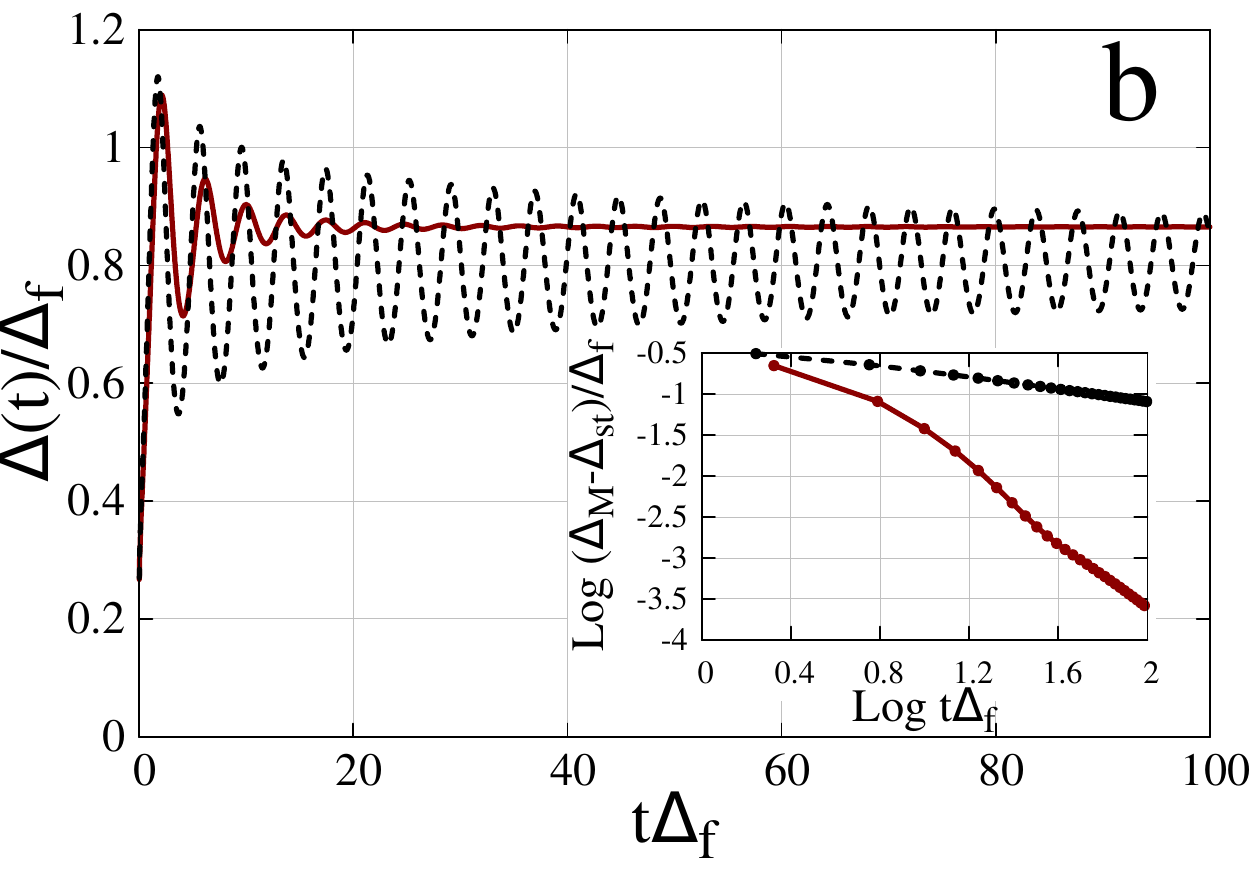}\\
 \includegraphics[width=40mm]{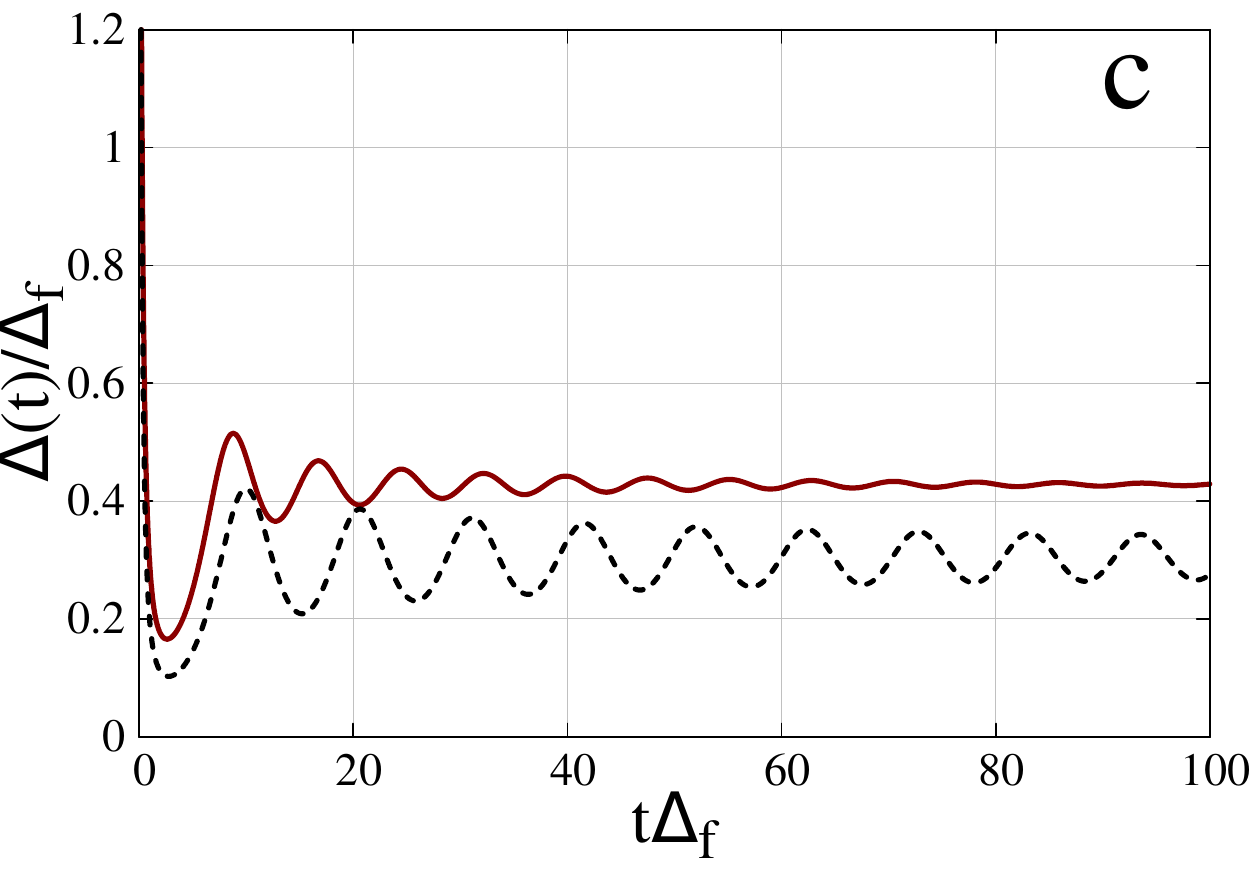}&
 \includegraphics[width=40mm]{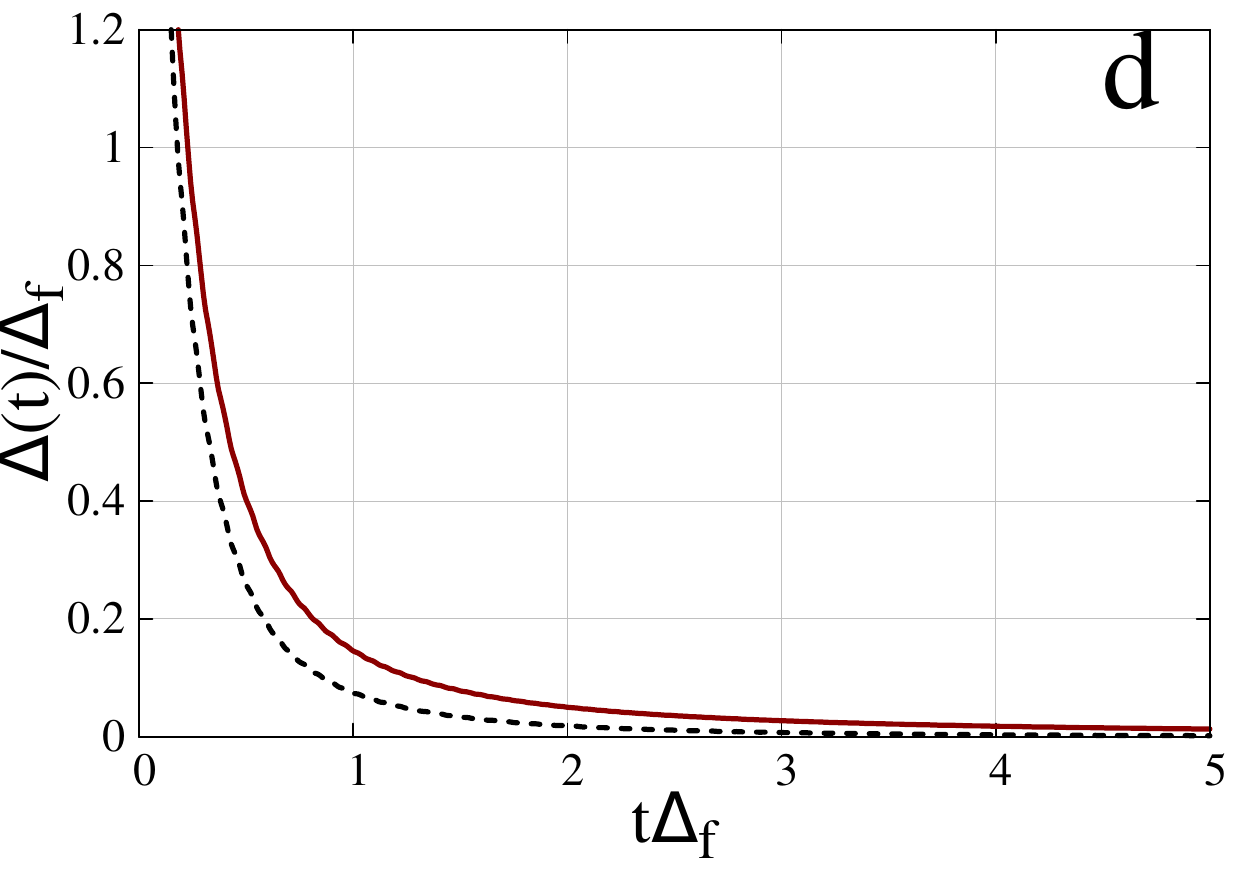}\\
 \end{tabular}
 \caption{(Color online) Plot of the gap dynamics for the $d$-wave (full red line) and $s$-wave (dashed black line) symmetries and for different quench parameters $\Delta_i/\Delta_f$: (a) $0.001$, (b) $0.2$, (c) $4.0$, and (d) $5.2$. Inset of panel (b): log-log plot of the local maxima of the gap.}
 \label{fig:fig1}
 \end{center}
 \end{figure}

 In Fig.~\ref{fig:fig1} we plot the time-dependent order parameter $\Delta(t)$ for $s$- and $d$-wave symmetry and for four different values of the quench parameter $\Delta_i/\Delta_f$, ranging from very small to very large ratios.
 In the $s$-wave case we recover three different dynamical regimes, in accordance with previous studies~\cite{BarankovLevitovPRL06}: For small $\Delta_i/\Delta_f$ the order parameter exhibits persistent oscillations between two limiting values; for intermediate $\Delta_i/\Delta_f$ it has damped oscillations towards a non-equilibrium asymptotic value $\Delta_\mathrm{st}\neq0$; and at large $\Delta_i/\Delta_f$ it has an overdamped exponential decay to zero.

 In the $d$-wave case we find damped oscillations for every $\Delta_i/\Delta_f$ except for very large values.
 On the qualitative level, it is remarkable the disappearance of the regime with persistent oscillations [panel (a) of Fig. \ref{fig:fig1}] and, more importantly, the much faster decay of the gap oscillations as compared to the $s$-wave case [panels (b) and (c) of Fig. \ref{fig:fig1}].
 On the contrary, for very large $\Delta_i/\Delta_f$ the behavior is similar, with both $s$- and $d$-wave gap decaying exponentially to zero.
 Indeed, as pointed out in Ref.~\cite{BarankovLevitovPRL06}, this regime is similar to a complete switch-off of the interaction, a case in which the structure factor $\gamma_\bk$ has clearly little influence.

 The increased damping of the $d$-wave gap dynamics is a signature of the existence of low-energy excitations, as it can be understood at least in the case of small quenches $\Delta_i/\Delta_f\simeq1$ for which we can calculate the linear response theory variation $\delta\Delta(t)=\Delta(t)-\Delta_i$~\cite{SM_dynDwaveSC}
 \begin{equation}
 \label{eqn:dDelta}
 \delta\Delta(t)/\Delta_i\propto\sum_{\bk}\,\frac{\gamma_\bk^2\,\eps_\bk^2}{E_{\bk i}^3}\left(1-\cos 2E_{\bk i}t \right)
 \end{equation}
 where $E_{\bk i}=\sqrt{\eps_\bk^2+\gamma_\bk^2\Delta_i^2}$.
 The time-dependent contribution to Eq.~(\ref{eqn:dDelta}) is dominated at long times by the low-energy modes, as it is evident if we replace the sum over momenta with an energy integral and we change variables in order to introduce the superconducting density of states $\rho(E)=\sum_\bk\delta(E-E_\bk)$.
 For the $s$-wave superconductor the density of states has a sharp edge at $\Delta_i$ where it has a squared-root divergence. This leads to power-law damped oscillations with frequency $2\Delta_i$.
 The $d$-wave symmetry introduces a qualitative difference in the density of states, which diverges only logarithmically at $\Delta_i$ and has finite value for energies down to zero.
 This results in oscillations which damp much faster.

 \begin{figure}[t]
 \begin{center}
 \includegraphics[width=80mm]{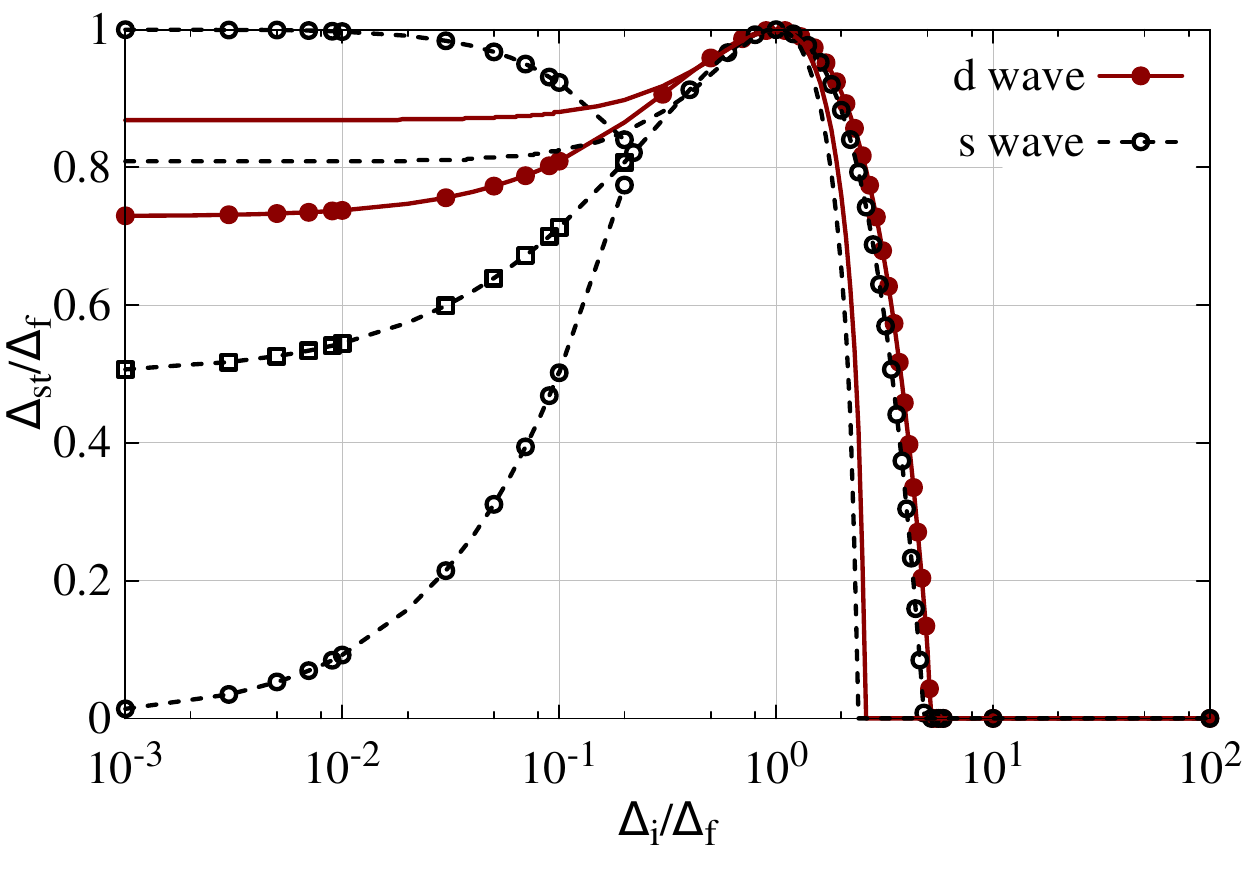}
 \caption{(Color online) Plot of the stationary gap $\Delta_\mathrm{st}/\Delta_f$ as a function of the quench parameter $\Delta_i/\Delta_f$ for the $d$-wave (full red line with full circles) and $s$-wave (dashed black line with empty circles) symmetries.
 For $\Delta_i/\Delta_f\lesssim0.2$ the $s$-wave gap exhibits undamped oscillations around the value indicated with a square, in this case the circles indicate the extrema of the oscillations.
 Curves without symbols: equilibrium gap $\Delta(T^*)$ corresponding to an effective temperature $T^*$.} 
 \label{fig:fig2}
 \end{center}
 \end{figure} 

\emph{Steady state. - } The different dynamical regimes and the long-time gap values for s- and $d$-wave symmetries are summarized in Fig.~\ref{fig:fig2} where, following Ref.~\cite{BarankovLevitovPRL06}, we plot $\Delta_\mathrm{st}$ as a function of the quench parameter $\Delta_i/\Delta_f$.
 For any value of $\Delta_i/\Delta_f$ we find that at long times the quench leads to a reduction of the gap with respect to the zero temperature equilibrium value $\Delta_f$.
 The difference between the s- and $d$-wave cases occurs for $\Delta_i/\Delta_f\lesssim0.2$, where the $d$-wave gap goes to a stationary value while the $s$-wave gap exhibits undamped oscillations~\cite{BarankovLevitovPRL06}.
 On the other hand, for $\Delta_i/\Delta_f\gtrsim0.2$ the gap reaches essentially the same asymptotic value, despite the much faster decay of the $d$-wave gap.
 
 It is important to emphasize that in the absence of pair-breaking scattering terms and of any real dissipation mechanism, the system persists in a nonequilibrium state.
In particular, the expectation values $\braket{c_{\bk\sigma}^\dagger(t)c_{\bk\sigma}(t)}$ and $\braket{c_{-\bk\down}(t)c_{\bk\up}(t)}$ do not come to a steady state and a stationary value of the gap is eventually reached only as a result of destructive interference between different momenta, a phenomenon which can also be interpreted in terms of a quench-induced decoherence~\cite{SchiroMitraPRL14}.

 For completeness, in Fig.~\ref{fig:fig2} we also plot $\Delta(T^*)$, the gap for a system in equilibrium at the temperature $T^*$ corresponding to the energy pumped into the system through the quench (lines without symbols).
 The system could eventually reach this thermal value if we include scattering processes not contained in the Hamiltonian [Eq.~(\ref{eqn:Hqp})].
 While the overall behavior of $\Delta(T^*)$ is qualitatively similar to $\Delta_\mathrm{st}$, the quantitative difference is substantial confirming the nonthermal character of the asymptotic stationary state.

\emph{Spectral features. - } We now focus on the spectral properties of the transient state in the $d$-wave case, thus moving a first step towards a comparison with recent time-resolved ARPES experiments on high-$T_c$ cuprates~\cite{GrafEtAlNatPhys11,SmallwoodEtAlScience12,SmallwoodEtAlPRB14}.
 These have tracked the evolution of the quasiparticle energy $E_\bk$ and weight $Z_\bk$, which are directly related to the lesser Green's function $G^<_\bk(t,t')=i\braket{c_{\bk\sigma}^\dagger(t)c_{\bk\sigma}(t')}$, whose Fourier transform at equilibrium and zero temperature has a peak at negative energy
 \begin{equation}
 -\frac{i}{\pi}G^<_{\bk\mathrm{eq}}(\omega)=Z_{\bk\mathrm{eq}}\delta(\omega+E_\bk)
  \label{eqn:G_eq}
 \end{equation}
 corresponding to the quasiparticle energy $E_\bk$ with a weight 
 \begin{equation}
 Z_{\bk\mathrm{eq}}=(1-\frac{\eps_\bk}{E_\bk}).
 \label{eqn:Z_eq}
 \end{equation}
 
 In principle, from the quasiparticle Hamiltonian [Eq.~(\ref{eqn:Hqp})] we can also calculate the out-of-equilibrium lesser Green's function.
 However, the presence of an arbitrary time-dependent order parameter $\Delta(t)$ and the fact that out of equilibrium the Green's function depends on both time arguments, makes this a rather challenging task that we leave for further studies.
 Here, in order to proceed analytically and obtain some physical insight on the main effect of the quench, we exploit the observation of the extremely fast dynamics of the $d$-wave gap.
 Hence, we assume that the quasiparticle modes do not have enough time to rearrange and approximate the actual dynamics with a sudden change of the gap $\Delta(t)=\Delta_i\theta(-t)+\Delta_\mathrm{st}\theta(t)$.

 This approximation is also based on some experimental results.
 In particular Refs.~\cite{SmallwoodEtAlScience12,SmallwoodEtAlPRB14} have highlighted two clearly distinct time scales for the gap dynamics following the pump pulse: on a short interval of approximately $0.3\,\mathrm{ps}$ the gap reaches a value smaller than at equilibrium ($\Delta_\mathrm{st}$ in our model) and on a longer time it relaxes to the equilibrium value, typically attained after $10-20\,\mathrm{ps}$.
 Our sudden approximation for the gap dynamics should therefore be reasonable for times of a few picoseconds after the pump pulse.
It is interesting to notice that on these time scales an effective temperature picture of the ARPES spectra is not adequate, as firmly pointed out in Ref.~\cite{SmallwoodEtAlPRB14}. Finally, if we consider the maximum gap value at equilibrium for Bi$2212$ $\Delta_f=60\mathrm{meV}$ we can estimate a time scale of about $0.01\mathrm{ps}$ for Fig.~\ref{fig:fig1}.

 Within this sudden approximation for the gap dynamics, we derive an analytical expression for $G^<_\bk(t,t')$ which is convenient to Fourier transform with respect to the time difference $t-t'$ and average over the waiting time $t'$ to finally obtain~\cite{SM_dynDwaveSC}
 \begin{equation}
 -i/\pi G^<_{\bk\mathrm{neq}}(\omega)=Z_{\bk\mathrm{neq}}^-\,\delta(\omega+E_{\bk})+Z_{\bk\mathrm{neq}}^+\delta(\omega-E_{\bk}),
 \end{equation}
 where $E_\bk= \sqrt{\eps_\bk^2+\gamma_\bk^2\Delta_\mathrm{st}^2}$ and $Z^\pm_{\bk\rm neq}$ are the out of equilibrium positive and negative energy weights.
 We notice that as a result of the sudden excitation also the positive energy peak has a finite occupation.
 In the following we will focus on the negative energy peak, which is the one observed in ARPES, and compare its weight $Z^-_{\bk \rm neq}$ to the equilibrium case. To this extent it is important to discuss first the interpretation of the quench protocol in the framework of pump-probe experiments.
 In the standard picture, the quench is used to describe the change of a Hamiltonian parameter, in this case $J(t)=J_i\theta(-t)+J_f\theta(t)$, as a result of an external perturbation.
 However this scenario --which can be realized in cold-atom systems-- is not directly relevant to solid-state experiments.
 In this context the Hamiltonian parameters can be considered largely independent of the excitation process and the quench is merely a theoretical tool to study the evolution of an out-of-equilibrium state.
 In this approach $J_i$ and $\Delta_i$ are just used to parameterize the initial state which results from the impulsive excitation, whereas the interaction parameter $J_f$ which controls the time evolution has to coincide with the actual interaction that characterizes the material.

 In this light, it is appropriate to compare the out-of-equilibrium spectral weight $Z_{\bk \rm neq}^-$ with the one at equilibrium [Eq.~(\ref{eqn:Z_eq})] with $\Delta_f$.
 The result of this calculation reads~\cite{SM_dynDwaveSC}
 \begin{equation}
 \label{eqn:Z_neq}
 \frac{Z_{\bk\mathrm{neq}}^-}{Z_{\bk \rm eq}}= \frac{1-\eps_{\bk}/E_{\bk}}{1-\eps_{\bk}/E_{\bk \rm f}}
 \left(\frac{1}{2}+\frac{\eps^2_{\bk}+\Delta_{\rm st}\Delta_{\rm i}\gamma_{\bk}^2}{2 E_{\bk}
 E_{\bk \rm i}}\right)
 \end{equation}
 which takes a particularly clear and interesting form if we expand in the neighborhood of the nodal lines, i.e., for $\gamma_{\bk}\rightarrow 0$.
 In this case we obtain
 $Z_{\bk\mathrm{neq}}^-/Z_{\bk\mathrm{eq}}=\left(\Delta_\mathrm{st}/\Delta_f\right)^2<1$
 for $\eps_{\bk}>0$ while $Z_{\bk\mathrm{neq}}^-/Z_{\bk\mathrm{eq}}=1-\alpha\gamma_{\bk}^2/4\eps_{\bk}^2$ for $\eps_{\bk}<0$, with $\alpha=\Delta_{\rm i}^2-\Delta_{\rm f}^2+2\Delta_{\rm st}^2-2\Delta_{\rm st}\Delta_{\rm i}>0$.
 In other words the nodal spectral weight is always reduced with respect to the equilibrium value, except possibly in a small region close to the Fermi surface.
 This reduction is clearly a nonthermal effect since finite temperature excitations of quasiparticle would lead to a reduction proportional to the Fermi function.

\emph{Conclusions. -} We have studied the real-time dynamics of a simple model of $d$-wave superconductor excited by a sudden perturbation.
 For every value of the quench parameter $\Delta_i/\Delta_f$ the system relaxes to a non-thermal stationary state with a gap parameter smaller than at equilibrium.
 The presence of nodal gapless excitations results in a much faster dynamics compared to the uniform $s$-wave case and to the disappearance of the regime of undamped oscillations which characterizes an isotropic superconductor when the quench parameter is smaller than 0.2.
 
  We have derived an analytical expression for the momentum-dependent photoemission spectral weight, which demonstrates a strong dependence on the momentum.
 In particular we have found a strong suppression of the weight at the gap nodes which is clearly of nonthermal nature, a result which is consistent with recent time-resolved ARPES experiments.

 M. C. and F. P. acknowledge financial support from European Research Council through the Starting Grant SUPERBAD (Grant Agreement 240524). We acknowledge discussions with A. Amaricci, D. Fausti, and C. Giannetti.

\bibliography{biblio}

\end{document}


\title{Supplementary Material for "Transient Dynamics of d-wave Superconductors After Sudden Excitations"}
\author{Francesco Peronaci}
\affiliation{International School for Advanced Studies (SISSA/ISAS) and CNR-IOM Democritos, Via Bonomea 265, 34136, Trieste, Italy}
\author{Marco Schir\'o}
\affiliation{Institut de Physique Th\'{e}orique, Universit\'{e} Paris Saclay, CNRS, CEA, F-91191 Gif-sur-Yvette, France}
\author{Massimo Capone}
\affiliation{International School for Advanced Studies (SISSA/ISAS) and CNR-IOM Democritos, Via Bonomea 265, 34136, Trieste, Italy}
\date{\today}

\maketitle

\section{Linear Response Regime and Enhanced Damping of a dwave Superconductor}

\begin{figure}[b]
 \begin{center}
 \includegraphics[width=80mm]{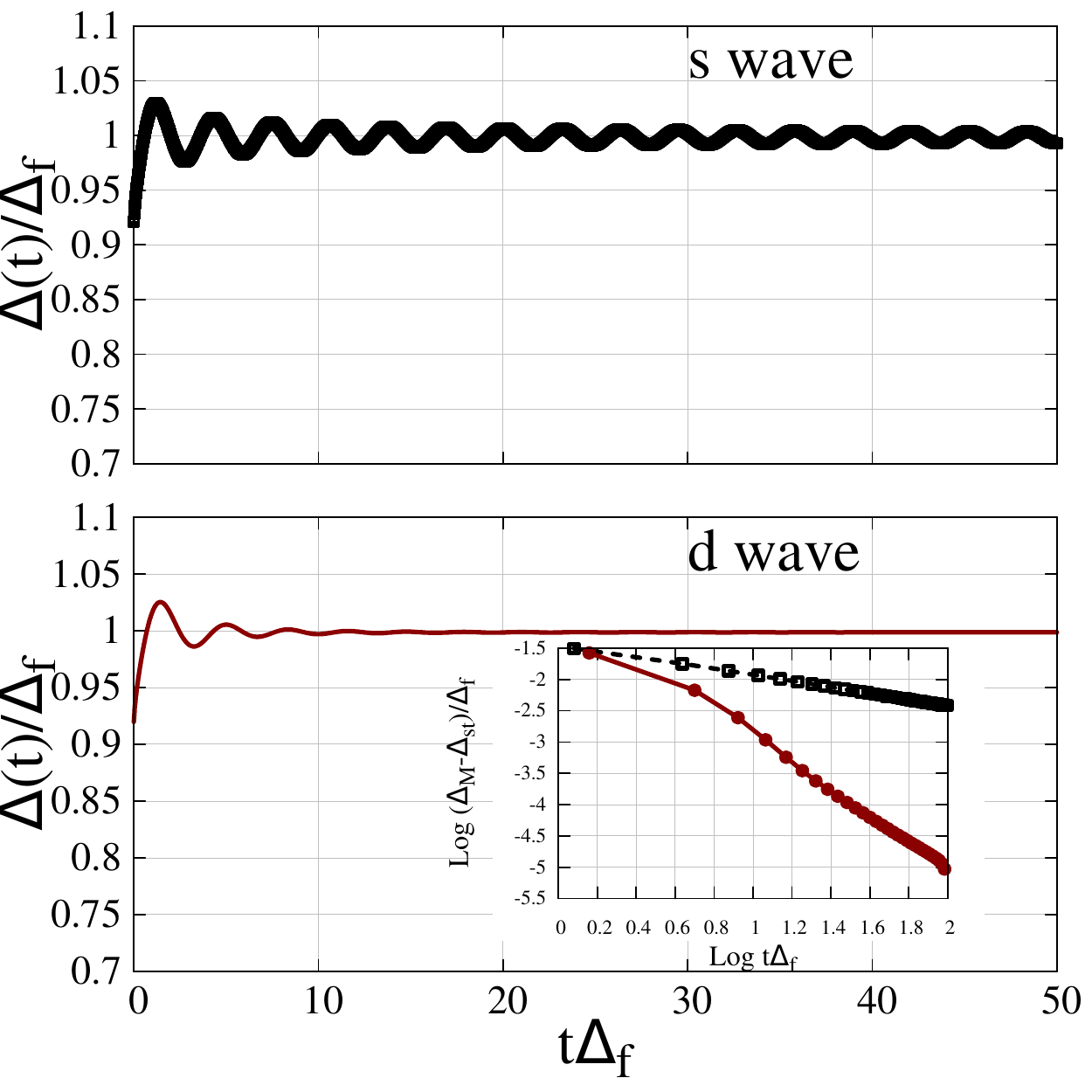}
 \includegraphics[width=80mm]{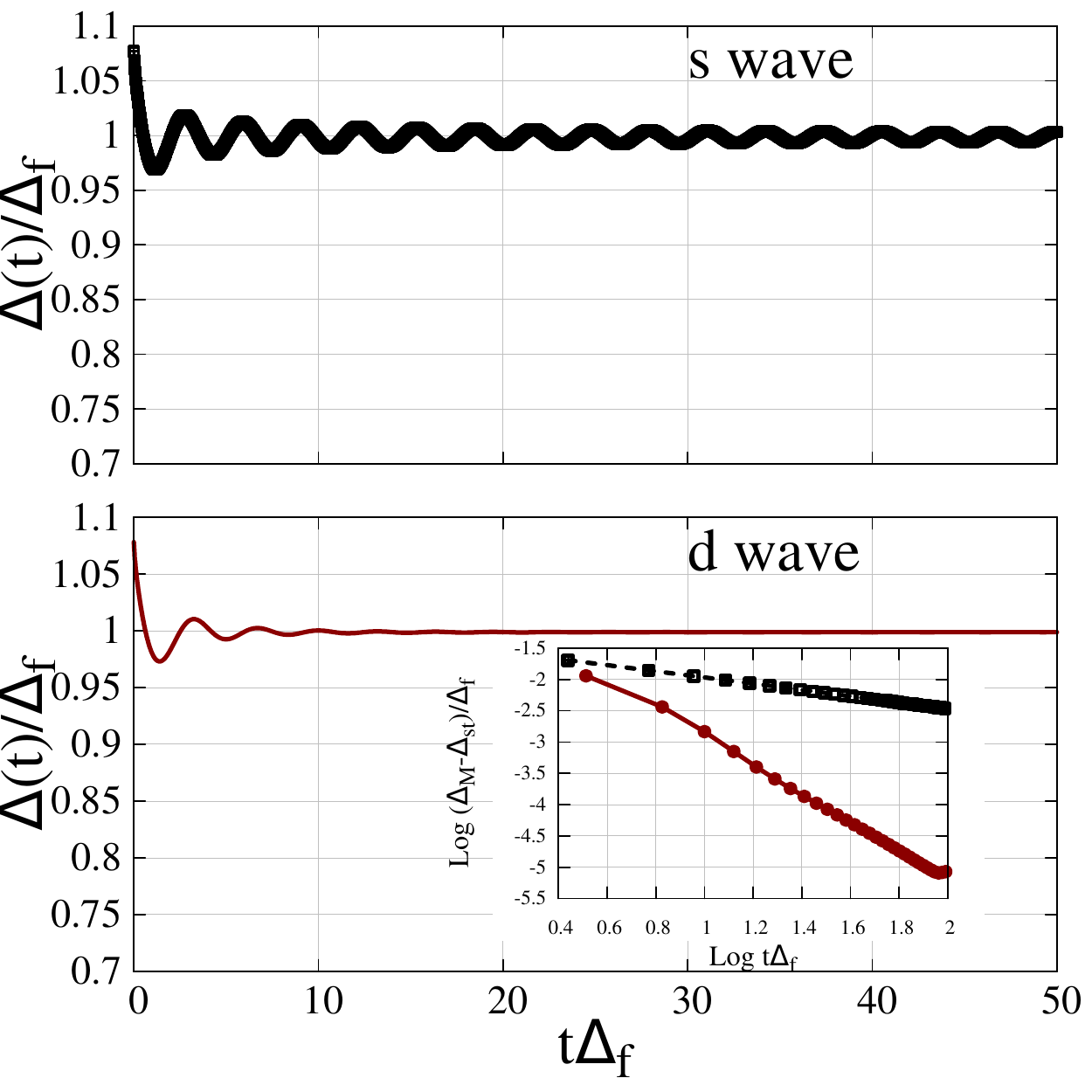}
 \caption{(Color online) Plot of the gap dynamics for d-wave (red, bottom panels) and s-wave (black, top panels) for different quench parameters $\Delta_\mathrm{i}/\Delta_\mathrm{f}$: $0.9$ (left panels) and $1.1$ (right panels). Inset: plot in log-log scale of the local maxima of the gap.}
 \label{fig:fig2SM}
 \end{center}
 \end{figure} 

We start from the time-dependent BCS Hamiltonian 
\be
H(t)=\sum_{\bk\sigma}\eps_{\bk}\,c^{\dagger}_{\bk\sigma}\,c_{\bk\sigma} -
\sum_{\bk\,\bp}\,J_{\bk\bp}(t)\,c^{\dagger}_{\bk\up}\,c^{\dagger}_{-\bk\downarrow}
c_{-\bp\downarrow}\,c_{\bp\up}
\ee
 for a sudden quench of the pairing interaction $J_{\bk\bp}(t)=J(t)\gamma_{\bk}\gamma_{\bp}$ where $J(t)=J_i\theta(-t)+J_f\theta(t)$ and with $\gamma_{\bk}=\left(\cos k_x-\cos k_y\right)$ for d-wave and $\gamma_{\bk}=1$ for s-wave.
The normal and anomalous Green's functions on the Keldysh contour are defined as
\bea
G_{\bk}(t,t')=-i\langle T_{\m{K}}c_{\bk\uparrow}(t) c^{\dagger}_{\bk\uparrow}(t') \rangle\\
\mathcal{G}_{\bk}(t,t')=-i\langle T_{\m{K}}c^{\dagger}_{-\bk\downarrow}(t)c_{-\bk\downarrow}(t) \rangle\\
F_{\bk}(t,t')=-i\langle T_{\m{K}}c_{\bk\uparrow}(t) c_{-\bk\downarrow}(t') \rangle
\eea
 where $T_{\m{K}}$ is the Keldysh time-ordering, from which we can obtain the time-dependent 
superconducting (SC) order parameter  $\Delta_{\bk}(t)=\langle c_{-\bk\downarrow}(t)c_{\bk\uparrow}(t)\rangle$
as
\be\label{eqn:Delta_t}
\Delta_{\bk}(t)=-iF^<_{\bk}(t,t) 
\ee
We now compute the anomalous Green's function within linear response in the strenght of the quench
 \be
 \delta \hat{V}(t>0)= \delta J \sum_{\bk\,\bp}\gamma_{\bk}\gamma_{\bp}\,c^{\dagger}_{\bk\up}\,c^{\dagger}_{-\bk\downarrow}
c_{-\bp\downarrow}\,c_{\bp\up}
 \ee 
 where $\delta J=J_i-J_f$, around the initial BCS ground-state with $J=J_i$. We obtain to first order
\be
\delta F_{\bk}(t,t') =F_{\bk}(t,t')-F^0_{\bk}(t,t')=
\sum_{\bp\bp'}\delta J_{\bp\bp'} \int_{\m{K}} dt_1 \langle \m{T}_{\m{K}}  
c_{\bk\uparrow}(t)c_{-\bk\downarrow}(t')c^{\dagger}_{\bp\up}(t_1)\,c^{\dagger}_{-\bp\downarrow}(t_1)
c_{-\bp'\downarrow}(t_1)\,c_{\bp'\up}(t_1)\rangle
\ee
We can evaluate the above expression within a BCS approximation, where we decouple normal and anomalous Green's functions, to get
\be
\delta F_{\bk}(t,t') =-\delta J\gamma_{\bk}\Delta_i \int_{\m{K}}dt_1\left[ G_{\bk}(t,t_1)\m{G}_{\bk}(t_1,t')+F_{\bk}(t,t_1)F_{\bk}(t_1,t')\right]
\ee 
where $\Delta_i$ is the gap of the initial BCS state. Using Eq.~(\ref{eqn:Delta_t}) and summing the order parameter over $\bk$ we get for $\delta \Delta(t)=\sum_{\bk}\gamma_{\bk}\,\delta\Delta_{\bk}(t)$ two contributions
\bea\label{eqn:dDelta}
\delta \Delta(t)/\Delta_i=\delta J \sum_{\bk}\,\gamma^2_{\bk} \int_0^t dt_1\,
i\left[G^R_{\bk}(t-t_1)\m{G}^<_{\bk}(t_1-t)+G^<_{\bk}(t-t_1)\m{G}^A_{\bk}(t_1-t)\right]+\\
+\delta J \sum_{\bk}\,\gamma^2_{\bk} \int_0^t dt_1\,i\left[
F^R_{\bk}(t-t_1) F^<_{\bk}(t_1-t)+F^<_{\bk}(t-t_1) F^A_{\bk}(t_1-t)\right]
\eea
The normal and anomalous retarded Green's functions are readily evaluated in real-time 
\bea\label{eqn:GF}
G^R_{\bk}(t)=-i\theta(t)\langle \left\{c_{\bk\uparrow}(t),c^{\dagger}_{\bk\uparrow}(0)\right\}\rangle=
-i\theta(t)\left( \cos^2\theta_{\bk}\,e^{-iE_{\bk}t}+\sin^2\theta_{\bk}\,e^{iE_{\bk}t}\right)\\
F^R_{\bk}(t)=-i\theta(t)\langle \left\{c_{\bk\uparrow}(t),c_{-\bk\downarrow}(0)\right\}\rangle=
\theta(t)\sin 2\theta_{\bk}\sin E_{\bk}t
\eea
while the advanced components are obtained by sending $t\rightarrow -t$ and after complex conjugation.
The lesser Green's functions read
\bea
G^<_{\bk}(t)=i\langle c^{\dagger}_{\bk\uparrow}(0)c_{\bk\uparrow}(t)\rangle=
i \sin^2\theta_{\bk}e^{iE_{\bk} t}\\
\m{G}^<_{\bk}(t)=i\langle c_{-\bk\downarrow}(0) c^{\dagger}_{-\bk\downarrow}(t)\rangle=
i \cos^2\theta_{\bk}e^{iE_{\bk} t}\\
F^<_{\bk}(t)=i\langle  c_{-\bk\downarrow}(0) c_{\bk\uparrow}(t)\rangle=
-i\sin\theta_{\bk}\cos\theta_{\bk}e^{iE_{\bk} t}
\eea
where the angle $\theta_{\bk}$ is defined as usual in BCS theory
\bea
\cos^2\theta_{\bk}=\frac{1}{2}\left(1+\frac{\eps_{\bk}}{E_{\bk}}\right)\\
\sin^2\theta_{\bk}=\frac{1}{2}\left(1-\frac{\eps_{\bk}}{E_{\bk}}\right)
\eea
with the quasiparticle spectrum $E_{\bk}=\sqrt{\eps_{\bk}^2+\gamma_{\bk}^2\Delta_i^2}$. Plugging the above results into the expression~(\ref{eqn:dDelta}) we finally obtain
\be
 \delta \Delta(t)/\Delta_i=\delta J \sum_{\bk}\,\frac{\gamma^2_{\bk}}{2E_{\bk}}\cos 2\theta_{\bk}
 \left[
 \cos^2\theta_{\bk}\left(e^{-i2E_{\bk}t}-1\right)-
 \sin^2\theta_{\bk}\left(e^{-i2E_{\bk}t}-1\right)
 \right]
\ee
The imaginary part of this expression
\be
\mbox{Im} \delta \Delta(t)/\Delta_i=-\delta J \sum_{\bk}\,\frac{\gamma^2_{\bk}}{2E_{\bk}}\cos 2\theta_{\bk}\sin 2E_{\bk}t=-\delta J \sum_{\bk}\,\frac{\gamma^2_{\bk}\eps_{\bk}}{2E^2_{\bk}}\sin 2E_{\bk}t=0
\ee
vanishes identically so that the final result is purely real and reads
\be
 \delta \Delta(t)/\Delta_i=-\delta J \sum_{\bk}\,\frac{\gamma^2_{\bk}\eps_{\bk}^2}{2E^3_{\bk}}
 \left(1-\cos 2E_{\bk}t \right)
\ee
In the d-wave case we can approximate the dispersion close to the nodes by its conelike structure, 
$E_{\bk}\simeq \sqrt{v_F^2k_1^2+v_{\Delta}^2k_2^2}$ and using polar coordinates $(E,\theta)$ to express $\eps^2_{\bk}=E^2\cos^2\theta$ and $\gamma_{\bk}^2=E^2\sin^2\theta/\Delta_i^2$ we get
\be
\delta \Delta_d(t)/\Delta_i=\delta \Delta_{d\infty}/\Delta_i+\frac{\pi\delta J}{8v_Fv_{\Delta}\Delta_i^2} 
 \int_0^{\Lambda} dEE^2\cos 2Et\simeq \delta \Delta_{d\infty}/\Delta_i + A_d/t
\\\ee
In the s-wave case we can insert the SC density of states $\rho(E)=\rho_0\theta(E-\Delta_i) E/\sqrt{E^2-\Delta_i^2}$ which has a sharp edge at $E=\Delta_i$, to get for the approach to the steady state
\bea
 \delta \Delta_s(t)/\Delta_i=\delta \Delta_{s\infty}/\Delta_i+\frac{\delta J\rho_0}{2}\int_{\Delta_i}^{\Lambda} dE\frac{\sqrt{E^2-\Delta_i^2}}{E^2}\cos 2Et\simeq \delta \Delta_{s\infty}/\Delta_i + A_s\cos (2\Delta_i t+\phi)/t^{\alpha}
\eea
The behavior of the time-dependent gap $\Delta(t)$ for a small amplitude quench, $\Delta_i/\Delta_f=0.9, 1.1$ is plotted for reference in Figure~\ref{fig:fig2SM}. We see that for a uniform s-wave gap, small coherent oscillations persist at long times while they are quenched on much shorter time scales in the d-wave case. The inset in both figures show the decay of the time-dependent envelope of the gap function, which confirms a faster power law decay due to the momentum anisotropy of the d-wave gap.

\section{Sudden Quench Approximation}

In the limit of a very fast order parameter dynamics, we can assume the quasiparticle (qp) Hamiltonian to experience an almost sudden change of the pairing field $\Delta_{\bk}(t)=\Delta(t)\gamma_{\bk}$, with $\Delta(t)=\Delta_i\theta(-t)+\Delta_{st}\theta(t)$. The problem can be than solved analytically by a time-dependent Bogolubov transformation.
We start by diagonalizing the initial qp Hamiltonian, that in the Nambu formalism $\bar\Psi_{\bk}=\left(c^{\dagger}_{\bk\uparrow}\;\;c_{-\bk\downarrow}\right)$ reads
\be
H_{qp}(\Delta_i)= \sum_{\bk\sigma}\bar{\Psi}_{\bk}\left(\eps_{\bk}\tau_z+\Delta_i\gamma_{\bk}\tau_x\right)\Psi_{\bk}
\ee
in terms of a new set of fermions
\be
\Phi_{\bk}=\exp\,\left(i\theta_{\bk i}\tau_y\right)\,\Psi_{\bk} 
\ee
such that
\be
H_{qp}(\Delta_i)= \sum_{\bk}\bar{\Phi}_{\bk}\,E_{\bk i}\tau_z\Phi_{\bk}
\ee
with $E_{\bk i}=\sqrt{\eps_{\bk}^2+\Delta_{\bk i}^2}$. Similarly for the final qp Hamiltonian,
\be 
H_{qp}(\Delta_{st})= \sum_{\bk\sigma}\bar{\Psi}_{\bk}\left(\eps_{\bk}\tau_z+\Delta_{st}\gamma_{\bk}\tau_x\right)\Psi_{\bk}
\ee
we can introduce a new set of fermions
\be
\Upsilon_{\bk}=\exp\,\left(i\theta_{\bk}\tau_y\right)\,\Psi_{\bk} 
\ee
such that
\be
H_{qp}(\Delta_{st})= \sum_{\bk}\bar{\Upsilon}_{\bk}\,E_{\bk}\tau_z\Upsilon_{\bk}
\ee
with $E_{\bk st}=\sqrt{\eps_{\bk}^2+\Delta_{\bk st}^2}$. Then, we can write the time evolution of the original fermionic field as
\be
\Psi_{\bk}(t)=
M_{\bk}(t)
\Phi_{\bk}(0)
\ee 
with 
$$
M_{\bk}(t)=
\left(
\begin{array}{ll}
f_{\bk}(t) & g_{\bk}(t)\\
-g_{\bk}^*(t) & f_{\bk}(t)^*
\end{array}
\right) 
$$
and
\bea
f_{\bk}(t)=\cos E_{\bk}t\,\cos \theta_{\bk i} - i\sin E_{\bk }t\,\cos\left(2\theta_{\bk }-\theta_{\bk i}\right)\\
g_{\bk}(t)=-\cos E_{\bk}t\,\sin\theta_{\bk i}-i\sin E_{\bk }t \sin\left(2\theta_{\bk }-\theta_{\bk i}\right)
\eea
We can then define the following Nambu Green's function (Gf) on the Keldysh contour, involving both normal (diagonal) and anomalous (off-diagonal) components
\be
G_{\bk\alpha\beta}(t,t')=-i\langle T_K\,\Psi_{\bk\alpha}(t)\bar{\Psi}_{\bk\beta}(t')\rangle 
\ee
and focus in particular on the lesser Gf, $G^<_{\bk\alpha\beta}(t,t')=i\langle\,\bar{\Psi}_{\bk\beta}(t')\Psi_{\bk\alpha}(t)\rangle$ whose normal component ($\alpha=\beta=1$)
\be 
G^<_{\bk}(t,t')=i\langle\,c^{\dagger}_{\bk\uparrow}(t')c_{\bk\uparrow}(t)\rangle
\ee
 is related to the ARPES signal, which typically probes the matrix element for transition from occupied to empty states. A simple algebra gives the result
\be\label{eqn:Gless}
G^<_{\bk}(t,t')=i\left[n(E_{\bk i})\,f_{\bk}(t)f^*_{\bk}(t')+(1-n(E_{\bk i}))g_{\bk}(t)g^*_{\bk}(t')\right]
\ee
where we have introduced the Fermi function $n(E)=1/(e^{\beta E}+1)$ with $\beta=1/T$ the inverse temperature of the initial state. It is easy to see that in equilibrium, i.e. no quench of the pairing field $\Delta_i=\Delta_f$, we recover the standard result 
\be\label{eqn:Gless_eq}
G^<_{\bk}(t-t')=i\left[n(E_{\bk i})\,\cos^2\theta_{\bk i}\,e^{-iE_{\bk i}(t-t')}+(1-n(E_{\bk i}))
\sin^2\theta_{\bk i}\,e^{iE_{\bk i}(t-t')}\right]
\ee
which in Fourier space gives
\be
G^{<}_{\bk}(\omega)=i\left[n(E_{\bk i})\,\cos^2\theta_{\bk i}\,\delta(\omega-E_{\bk i})+(1-n(E_{\bk i}))
\sin^2\theta_{\bk i}\,\delta(\omega+E_{\bk i})\right]
\ee
At low enough temperature, $n(E_{\bk i})\simeq 0$ and only the negative energy peak is present and this will be the one whose evolution with the quench amplitude we will be following. In the non-equilibrium quenched case the lesser Gf will depend in general from both $t-t'\equiv\tau$ and $t+t'$. 
It is convenient to use $\tau$ and $t$ as independent variables and Fourier transform with respect to the former
\be 
G^<_{\bk}(\omega,t)=\int d\tau e^{i\omega \tau}\,G^<_{\bk}(\tau,2t-\tau)=
i\left[\left(A_{\bk}+\frac{C_{\bk}}{2}e^{-2iE_{\bk }t}\right)\delta(\omega+E_{\bk })+
\left(B_{\bk}+\frac{C_{\bk}}{2}e^{2iE_{\bk }t}\right)
\delta(\omega-E_{\bk })\right]
\ee
\begin{figure}[t]
 \begin{center}
 \includegraphics[width=80mm]{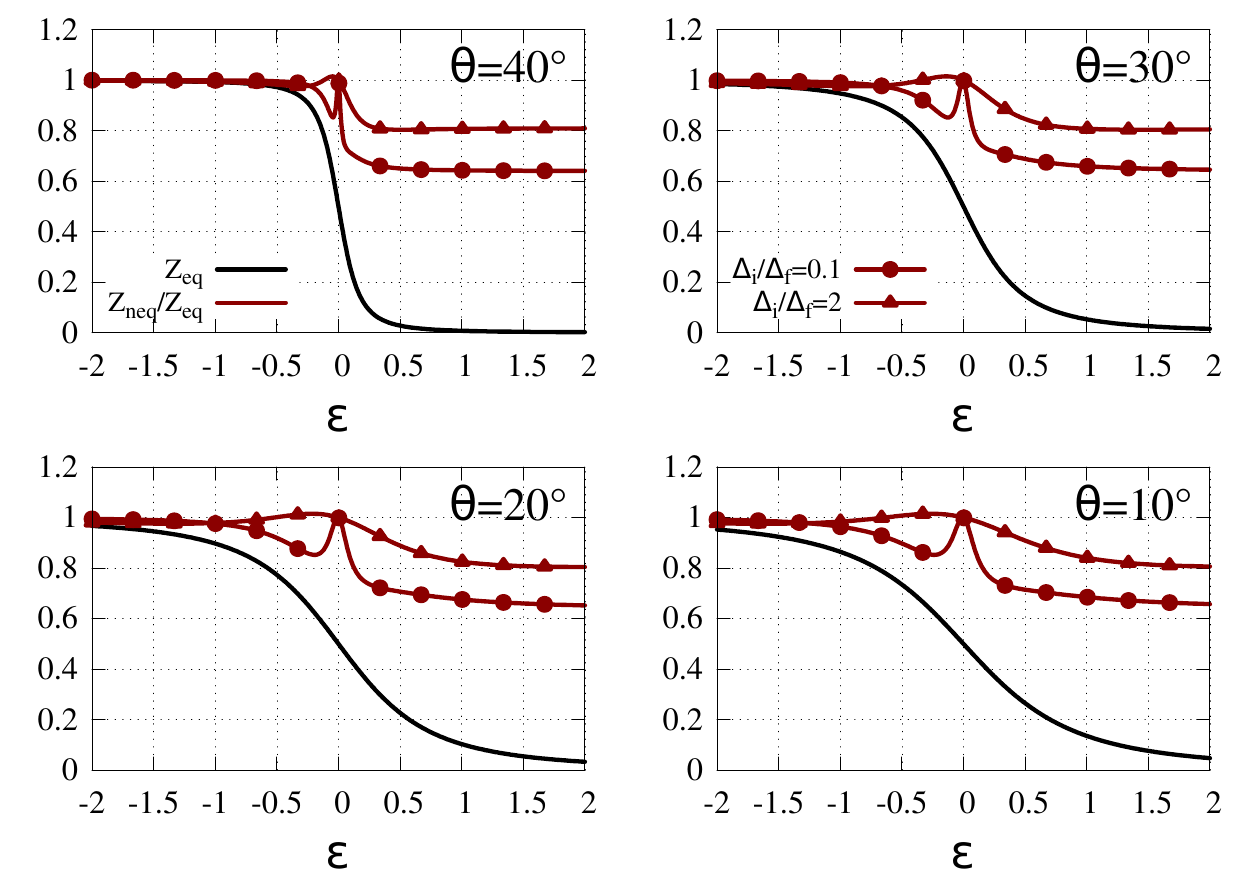}
 \caption{Plot of $Z_{\bk eq}$ and of the ratio $Z^-_{\bk neq}/Z_{\bk eq}$ as a function
 of $\eps_{\bk}$ for different polar angle $\theta$ in k-space and for two different
 quench parameter $\Delta_i/\Delta_f$.}
 \label{fig:fig1SM}
 \end{center}
 \end{figure} 
with respectively
\bea
A_{\bk}= \frac{1}{4}n(E_{\bk i})\left(\cos\theta_{\bk i}-\cos (2\theta_{\bk }-\theta_{\bk i})\right)^2+
\frac{1}{4}(1-n(E_{\bk i}))\left(\sin\theta_{\bk i}+\sin (2\theta_{\bk }-\theta_{\bk i})\right)^2\\
B_{\bk}= \frac{1}{4}n(E_{\bk i})\left(\cos\theta_{\bk i}+\cos (2\theta_{\bk }-\theta_{\bk i})\right)^2+
\frac{1}{4}(1-n(E_{\bk i}))\left(\sin\theta_{\bk i}-\sin (2\theta_{\bk }-\theta_{\bk i})\right)^2\\
C_{\bk}=\frac{1}{2}n(E_{\bk i})\left(\cos^2\theta_{\bk i}-\cos^2 (2\theta_{\bk }-\theta_{\bk i})\right)+
\frac{1}{2}(1-n(E_{\bk i}))\left(\sin^2\theta_{\bk i}-\sin^2 (2\theta_{\bk }-\theta_{\bk i})\right)
\eea
While it is tempting to identify the coefficients in front of the two delta functions as time-dependent spectral weight 
they are not positive definite unless we average over the waiting time $t$ 
\be 
G^<_{\bk}(\omega)=\mbox{lim}_{t\rightarrow
\infty}\frac{1}{t}\int_0^{t} dT\,G^<_{\bk}(\omega,T)
\ee
to finally obtain
\be
(-i/\pi)G^<_{\bk}(\omega) = Z^-_{\bk\rm neq}\,\delta(\omega+E_{\bk })+Z^+_{\bk\rm neq} \delta(\omega-E_{\bk })
\ee
where, respectively, $Z^-_{\bk\rm neq}=A_{\bk}$ and  $Z^+_{\bk\rm neq}=B_{\bk}$. We can now look at the spectral weight associated to the negative frequency peak and compare it with the equilibrium one, for a final gap $\Delta_{f}$. The result reads
\be\label{eqn:ratio}
\frac{Z_{\bk\mathrm{neq}}^-}{Z_{\bk \rm eq}}= \left(\frac{1-\eps_{\bk}/E_{\bk}}{1-\eps_{\bk}/E_{\bk f}}\right)
\left(\frac{1}{2}+\frac{\eps^2_{\bk}+\Delta_{st}\Delta_i\gamma_{\bk}^2}{2 E_{\bk} E_{\bk i}}\right)
\ee
As we discussed in the main text, we can expand the dispersion close to the nodes, for $\gamma_{\bk}\Delta/\eps_{\bk}\ll 1$ to obtain an analytical expression
$$
\frac{Z_{\bk\mathrm{neq}}^-}{Z_{\bk \rm eq}}\simeq
\left\{
\begin{array}{lll}
(\Delta_{st}/\Delta_f)^2 &  & \eps_{\bk}>0\\
 & & \\
1-\alpha \gamma_{\bk}^2/4\eps_{\bk}^2 &  & \eps_{\bk}<0\\
 \end{array}
\right.
$$
where $\alpha=\Delta_i^2-\Delta_f^2+2\Delta_{st}^2-2\Delta_{st}\Delta_i $ turns out to be positive for all quench parameters. In both cases we get $\frac{Z_{\bk\mathrm{neq}}^-}{Z_{\bk \rm eq}}<1$ from which we conclude that, at least far from the Fermi surface the nodal spectral weight is reduced with respect to equilibrium. We plot the individual weights as well as their ratio~(\ref{eqn:ratio}) as a function of energy in figure~\ref{fig:fig1SM}, for different values of the quench parameter $\Delta_i/\Delta_f$ and for different polar directions in k-space. We notice that the ratio is lesser than one for most values of the quench parameter, unless $\Delta_i\gg\Delta_f$ and in a region close to the Fermi surface that shrinks approaching the nodal direction, thus confirming qualitatively the analytical result.